%% file: paperMain.tex
\documentclass[conference]{IEEEtran}
\IEEEoverridecommandlockouts
\usepackage{cite}
\usepackage{amsmath,amssymb,amsfonts}
\usepackage{algorithm}
\usepackage[noend]{algpseudocode}
\usepackage{graphicx}
\usepackage{textcomp}
\usepackage{xcolor}
\usepackage[normalem]{ulem}

\usepackage{url}

\input{macros}

\def\BibTeX{{\rm B\kern-.05em{\sc i\kern-.025em b}\kern-.08em
    T\kern-.1667em\lower.7ex\hbox{E}\kern-.125emX}}

\begin{document}

\title{\LARGE Automatic Generation of Hierarchical Contracts for Resilience in Cyber-Physical Systems}

\author{\IEEEauthorblockN{1\textsuperscript{st} Zhiheng Xu}
\IEEEauthorblockA{\textit{Nanyang Technological University}\\
Singapore \\
zhiheng.xu@ntu.edu.sg}
\and
\IEEEauthorblockN{2\textsuperscript{nd} Daniel Jun Xian Ng}
\IEEEauthorblockA{\textit{Nanyang Technological University}\\
Singapore \\
danielngjj@ntu.edu.sg}
\and
\IEEEauthorblockN{3\textsuperscript{rd} Arvind Easwaran}
\IEEEauthorblockA{\textit{Nanyang Technological University}\\
Singapore \\
arvinde@ntu.edu.sg}
}

\maketitle

\begin{abstract}
With the growing scale of Cyber-Physical Systems (CPSs), it is challenging to maintain their stability under all operating conditions. How to reduce the downtime and locate the failures becomes a core issue in system design. In this paper, we employ a hierarchical contract-based resilience framework to guarantee the stability of CPS. In this framework, we use Assume Guarantee (A-G) contracts to monitor the non-functional properties of individual components (e.g., power and latency), and hierarchically compose such contracts to deduce information about faults at the system level. The hierarchical contracts enable rapid fault detection in large-scale CPS. However, due to the vast number of components in CPS, manually designing numerous contracts and the hierarchy becomes challenging. To address this issue, we propose a technique to automatically decompose a root contract into multiple lower-level contracts depending on I/O dependencies between components. We then formulate a multi-objective optimization problem to search the optimal parameters of each lower-level contract. This enables automatic contract refinement taking into consideration the communication overhead between components. Finally, we use a case study from the manufacturing domain to experimentally demonstrate the benefits of the proposed framework.
\end{abstract}


\input{Introduction}

\input{SysModel}

\input{AutoGen}

\input{Simulation}

\input{Conclusions}

\clearpage

\section*{Acknowledgment}
This work was conducted within the Delta-NTU Corporate Lab for Cyber-Physical Systems with funding support from Delta Electronics Inc. and the National Research Foundation (NRF) Singapore under the Corporation Lab@University Scheme.

\vspace{-1mm}
\bibliographystyle{IEEEtran}
\bibliography{Bibliography}

\end{document}

%% file: macros.tex
\newcommand{\calA}{\ensuremath{\mathcal{A}}}

\newcommand{\calD}{\ensuremath{\mathcal{D}}}

\newcommand{\calU}{\ensuremath{\mathcal{U}}}

\newcommand{\calN}{\ensuremath{\mathcal{N}}}

\newcommand{\calG}{\ensuremath{\mathcal{G}}}

\newcommand{\calX}{\ensuremath{\mathcal{X}}}
\newcommand{\calY}{\ensuremath{\mathcal{Y}}}

\newcommand{\R}{\ensuremath{\mathbb{R}}}

\newcommand{\Z}{\ensuremath{\mathbb{Z}}}

\newcommand{\tx}{\ensuremath{\tilde{x}}}

\newcommand{\tsigma}{\ensuremath{\tilde{\sigma}}}

\newcommand{\tmu}{\ensuremath{\tilde{\mu}}}

\newcommand{\bx}{\ensuremath{\bar{x}}}

\newtheorem{remark}{Remark}

\newtheorem{definition}{Definition}
\newtheorem{proposition}{Proposition}
\newtheorem{contract}{Contract}
\newtheorem{component}{Component}
\newtheorem{theorem}{Theorem}

\makeatletter
\def\BState{\State\hskip-\ALG@thistlm}
\makeatother

%% file: Introduction.tex
\section{Introduction}

Under the Industry 4.0 initiative \cite{iFour}, conventional factories and infrastructures are evolving into ``smart systems'', which closely integrate the physical devices and equipment with the cyber-infrastructure (computation and communication). Recently, such large-scale CPS play an increasingly crucial role in various critical industries, such as intelligent transportation systems \cite{xiong2015cyber}, smart power grids \cite{mo2012cyber}, industrial manufacturing systems \cite{wang2015current}, etc. However, this rapid evolution of CPS has led to a significant increase in system complexity. This further introduces challenges in meeting all the system requirements during design and execution, particularly in the presence of systematic faults.

Due to the critical role of CPS, many recent studies have focused on the resiliency of such systems to various faults. However, the growing complexity and scale of these systems make it challenging to achieve this goal. When a fault occurs, it might require a significant amount of time as well as communication to identify the failure, diagnose the fault, and recover the system. Therefore, achieving rapid fault detection and diagnosis becomes a central issue in system design in the design of such systems.

To reduce the failure rates in CPS, researchers have used Non-Functional Properties (NFPs) to evaluate the Quality of Service (QoS) that the system can provide. In system engineering, a NFP is a requirement describing the criteria to judge the system's performance \cite{chung2012non}. Given the designers' requirements about the NFPs, one can design Assume-Guarantee (A-G) contracts as defined in~\cite{AG_Contracts}, to observe the NFPs of interest in a centralized manner. For instance, the NFPs of interest can be execution latency or power consumption. An A-G contract can be used to ensure these NFPs remain in viable ranges for the entire CPS. Otherwise, when the system violates the contract, an alarm will be sent to the operators. However, the main challenge of a large-scale CPS are its vast number of components. Whenever the system violates the contract, it is difficult to identify the source of the fault, i.e., this centralized solution is not sufficient to achieve rapid detection and diagnosis of the fault.

\begin{figure}[thp]
	\centering
	\includegraphics[width=8.5cm]{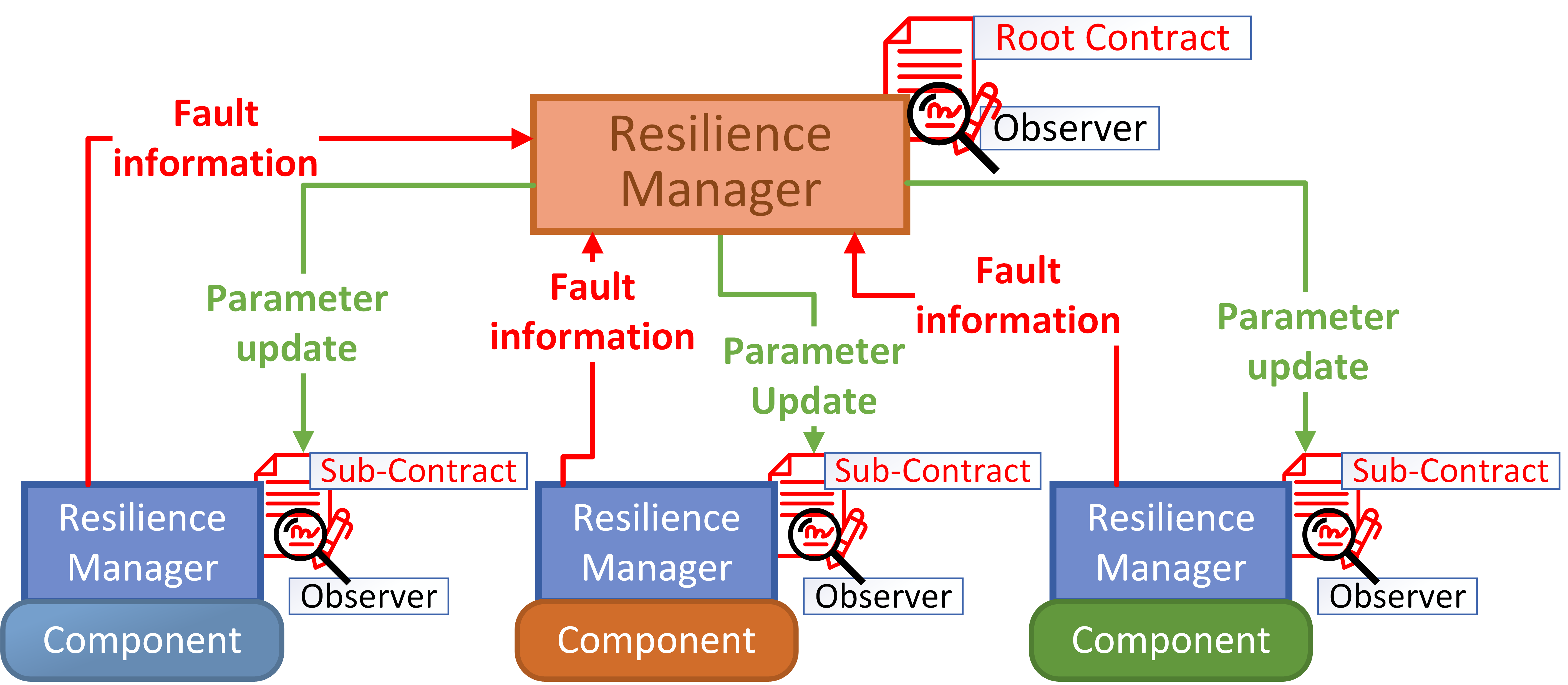}
	\caption{The Hierarchical Contract-Based Resilience Framework: The contract-based hierarchy has a root contract and multiple sub-contracts in each layer, and each RM manages a contract by using an observer to monitor the NFPs of the components.}\label{fig:hierFrame}

\end{figure}

To address the above issue, one can decompose the root contract (i.e., the original A-G contract) into multiple sub-contracts. Each sub-contract only monitors the NFPs of a specific component in the system. Hence, we can rapidly identify the source of the fault if we observe the violation of the corresponding sub-contract. However, such a simple and independent decomposition will make the CPS sensitive to random disturbances and false alarms. One false alarm in a component could impact the process of the entire system, increasing downtime and degrading system performance. For example, a jitter noise in a specific component may violate its sub-contract, shutting down the whole system when it could have resolved this jitter noise problem on its own. Further, such independent decomposition may not even be feasible in some cases, e.g., end-to-end latency requirements. Hence, a fully decentralized framework is also not always attractive.

To enhance resiliency as well as reduce false alarms, in our previous work, we have proposed a Hierarchical Contract-based Resilience Framework \cite{COMP_HContract}. As illustrated in Fig.~\ref{fig:hierFrame}, in this contract-based hierarchy, we have a root contract, which specifies the overall requirements of the system, and multiple sub-contracts, which present the specifications of each component. During the runtime of the CPS, software observers monitor the behaviors of the components. If any abnormal behavior violates a sub-contract, the observer will report to the root Resilience Manager (RM), indicating a fault. The root RM will verify whether the report is a false alarm by analyzing the overall information of the system available to it. In the contract-based hierarchy, the root contract monitors the overall NFP of the system, and the sub-contracts capture specific properties of individual components.

Constructing a two-level contract-based hierarchy has two important steps: (1) decomposing the root contract into various sub-contracts; (2) refining the sub-contracts. The decomposition ensures that the framework can isolate the faulty components quickly. While the refinement guarantees that the root RM has some amount of flexibility to resolve random disturbances or false alarms, thus reducing the downtime of the system.

Despite the advantages of the contract-based hierarchy, new challenges arise in a large-scale CPS with a vast number of components. It is challenging to decompose the root contract and refine numerous sub-contracts manually. Therefore, in this paper, we develop an algorithm to generate such hierarchical contracts automatically. As part of the automated solution, we propose a criterion to evaluate system performance based on the specific parameters of the contract and formulate an optimization problem to find the optimal settings. We summarize our main contributions as follows:
\vspace{1mm}
\begin{itemize}
  \item We develop an algorithm to achieve a simple contract decomposition. Using the algorithm, we can decompose a root contract into multiple sub-contracts based on the I/O information about the components and the root contract.\
  \item We design a criterion to evaluate the tradeoff between flexibility and communication cost of the contract-based hierarchy. Based on this criterion, we formulate an optimization problem to refine the sub-contracts.
      \vspace{1mm}
  \item We use dual decomposition to solve the optimization problem. The dual decomposition has a plug-and-play feature that allow the root system to add new component. Besides, we can use the dual approach to update the solutions efficiently whenever the system updates the parameters.
      \vspace{1mm}
  \item Based on the proposed algorithms, we develop a software tool suite to implement the automated algorithm for real applications. We use algorithms to generate contracts for a testbed.
      \vspace{1mm}
  \item Our experiments validate the implemented tool suite on a manufacturing case study and verify the performance of the system's resilience based on a simple hierarchical contract.
  %Talk about the case study
\end{itemize}

In our framework, even though we only study the case with one NFP, we can extend to the case study with multiple NFPs by designing multiple independent contract-based hierarchies. The proposed algorithms focus on the generation of a two-level hierarchy. However, we can also technically extend the work to multi-level hierarchies by running the algorithms iteratively.

To testify the algorithm, we use a Fischertechnik testbed which sorts tokens based on its color. In this testbed, there are three main components which performs their duties in a serial fashion. As such, there is an end-to-end execution latency requirement on the components for the token to be successfully sorted. We design a root contract and use the proposed algorithms to generate a valid contract-based hierarchy for the testbed.

\subsection{Related Work}

Many researchers have focused on contract-based design to enhance the reliability of industrial systems. Gauer et al. \cite{bauer2012moving} have presented the relationship between specification of components' behaviors and contracts. Alberto et al. \cite{Taming} have proposed a contract-based design to address critical issues such as variability, uncertainty, and life-cycle of a product family. Nuzzo et al. \cite{CHASE} have presented CHASE (Contract-based Heterogeneous Analysis and System Exploration) to achieve requirements capture, formalization and validation of CPS. The CHASE framework combines a front-end language with a verification back-end based on contracts.

For complex engineering systems, Filippidis et al. \cite{Layering} have proposed a decomposition algorithm to construct lower-level contracts for each component. The decomposition algorithm eliminates irrelevant variables to simplify the components' specifications. To decompose a root contract, Le et al. \cite{le2016contract} have presented conditions to verify the validity of the decomposition. They have developed algorithms to refine the sub-contracts to meet those conditions. However, the above work has not sufficiently addressed the issues of massive contract generation. Setting us apart from their work, we focus on the automatic generation of hierarchical contracts for CPS.

Our work also relates to the issue of resiliency in CPS. To maintain the stability and availability of the system, many researchers have focused on enhancing the resiliency of CPS. Pasqualetti et al. \cite{pasqualetti} have applied control-theoretical methods to strengthen the resiliency of CPS. In \cite{zhu}, Zhu et al. have analyzed trade-offs between robustness and resilience of modern industrial CPS. Through the analysis, they have proposed a hybrid theoretical framework to achieve robust and resilient control with an application to smart power systems. In our previous work \cite{xu2018cross}, we have developed a cross-layer approach to attain secure and resilient control of networked robotic systems. However, the above research cannot sufficiently capture the increasing complexity and variability of a large-scale CPS.

\subsection{Organization of the Paper}

We organize the remainder of the paper as follows. Section ~\ref{sec:SystemModel} presents the background and system model. Section~\ref{sec:2level} proposes the algorithms to achieve contract decomposition and refinement. Section~\ref{sec:experiment} illustrates the implementation of the proposed mechanism and the experimental results. Finally, Section~\ref{sec:conclude} concludes the paper.

%% file: SysModel.tex
\vspace{1mm}
\section{System Model and Problem Statement} \label{sec:SystemModel}

In this section, we first introduce our system model. Given the system model and the user's NFP requirements, we define a contract to guarantee the system's performance. We then formulate the problem to construct the contract-based hierarchy automatically.

\subsection{System Model and Contract Definition}

We assume that a CPS has a root system, denoted by $S_r$, which comprises all the components. Root system $S_r$ has $n$ number ($n$ is a positive integer) of components, denoted by $S_i$, for $i\in\calN:=\{1,\dots, n\}$. Each component $S_i$ has its inputs, outputs, and a Non-Functional Property (NFP) that determines its operational performance. We present a formal definition of the component's model as follows.

\begin{definition} (Component's Model)\label{def:comp}
  A component $S_i$ is a 3-tuple, i.e., $S_i := (u_i, y_i, x_i)$, where $u_i$ is the input, $y_i$ is the output, $x_i\in\calX_i$ is the estimated value of the NFP. Set $\calX_i$ specifies the feasible range of $x_i$.
\end{definition}

\begin{remark}
  In our framework, input $u_i$ and output $y_i$ can be any type of variables, e.g., real or boolean variables.
\end{remark}

In this paper, we present two feasible assumptions of the NFP's variable $x_i$: (1) $x_i\in\calX_i$ is a stochastic variable with a mean $\mu_i\in\R$ and a standard deviation $\sigma_i\in\R$; (2) the value of $x_i$ is summable, i.e., the total values of $x_i$ and $x_j$ is $x_i+x_j$, for $i,j\in\calN$. One typical example that satisfies these two assumptions is execution latency. It is straightforward to see that latency satisfies additivity, and Wilhelm et al. \cite{wilhelm2008worst} have shown that the execution time of physical devices follows a Gaussian distribution. Furthermore, we define that $\calX_i:=[a_i, b_i]$, where $a_i, b_i\in\R$ are the lower and upper bounds.
For example, let the NFP be the execution latency. Then, the latency of each components must stay in a feasible range with given lower and upper bounds.

\begin{figure}[thp]
	\centering
	\includegraphics[width=8cm]{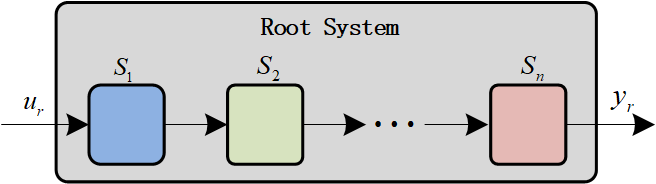}
	\caption{The architecture of the root system: all the components are connected in a serial structure.} \label{fig:rootSys}
\end{figure}

\begin{figure}[thp]
	\centering
	\includegraphics[width=8cm]{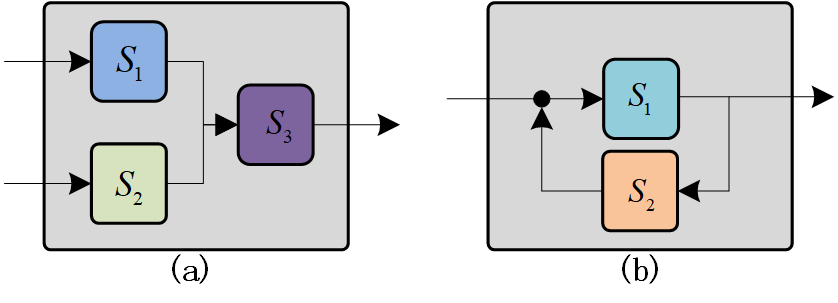}
    \vspace{-2mm}
    \caption{Possible structures in the system: (a) component $S_3$ depends on the
    outputs of components $S_1$ and $S_2$; (b) components $S_1$ and $S_2$ form a feedback structure.} \label{fig:compStructure}
\end{figure}

As illustrated in Fig.~\ref{fig:rootSys}, we assume $S_r$ connects all the components in a serial structure. However, in real applications, the CPSs can have various structures. As shown in Fig.~\ref{fig:compStructure} (a), components $S_1$ and $S_2$ can be in a parallel structure with the inputs of component $S_3$ depending on the outputs of $S_1$ and $S_2$. In this case, we can treat these three components as a new component, whose inputs and outputs are $\{u_1, u_2\}$ and $y_3$, respectively. Similarly, in Fig.~\ref{fig:compStructure} (b), we can treat the combination of $S_1$ and $S_2$ as a new component, which internally forms a feedback structure.

Note that root system $S_r$ is also a 3-tuple, i.e., $S_r := (u_r, y_r, x_r)$, where the input is $u_r:=u_1$, and the output is $y_r := y_n$. Inside the root system, all the components must satisfy the following conditions
\begin{eqnarray}
  y_i = u_{i+1}, \text{~for~} i\in\{1, \dots, n-1\}.\nonumber
\end{eqnarray}
Due to this serial structure, the overall performance of $S_r$ depends on the total NFP's values of all $ n $ components, i.e., $x_r := \sum_{i=1}^{n} x_i$.

To guarantee the performance of $S_r$, the designers will specify requirements on $x_r$, i.e., a threshold $\bx_r$ that $x_r$ should remain below. To monitor $x_r$, we introduce an Assume-Guarantee (A-G) root contract, defined as follows.

\begin{definition}(Root Contract)\label{def:contract}
A root contract is a 4-tuple, i.e., $C_r:= (u_r, y_r, \calA_r, \calG_r(\bx_r))$,
where $\calU_r$ and $\calY_r$ are the inputs and outputs of system; $\calA_r$ is the set of assumptions; $\calG_r(\bx_r)$ is the set of guarantees, defined by
\begin{eqnarray}
    \calG_{r}(\bx_{r}) := \biggl\{x_{r} = \sum_{i=1}^{n} x_i \ \biggl| \
  x_{r} \le \bx_{r} \biggl\}. \nonumber
\end{eqnarray}
\end{definition}

\begin{remark}
  According to Definition~\ref{def:contract}, we only use root contract $C_r$ to monitor one type of NFP. If the CPS has multiple NFPs, then we can design different independent contracts to monitor those NFPs.
\end{remark}

Based on the designers' requirements, we can design a root contract $C_r$ to monitor $x_r$. However, whenever $S_r$ violates $C_r$, it is challenging to identify which component is responsible for the failure. To solve this issue, we will introduce a contract-based hierarchy in the following subsection.

\subsection{Contract-Based Hierarchy}

Only using root contract $C_r$ hides important internal information about the faults. One solution is to decompose $C_r$ into sub-contracts $\{C_1, \dots, C_n\}$, i.e.,
\begin{eqnarray}
  C_r = C_1\otimes C_2 \otimes \cdots \otimes C_n,  \label{eq:decom1}
\end{eqnarray}
where $\otimes$ denotes the operator of the contract composition. We define the contract composition as follows.
\begin{definition}{Composition (\cite[Table IV]{AG_Contracts}):}\label{def:com}
	Given contracts $C_r$ $= ($ $u_r$, $y_r$, $\calA_r$, $\calG_r{(\bx_r)}$, $ )$ and $C_1$ $=($ $u_1$, $y_1$, $\calA_1, \calG_1(\bx_1))$, $\dots$, $C_n=( u_n, y_n, \calA_n, \calG_n(\bx_n))$, $C = C_1 \otimes C_2 \otimes \dots \otimes C_n$ iff the following conditions are satisfied:
	\begin{itemize}
		\item For $i=2,3,\dots,n$, $y_{i-1} :=  u_i$,
		\item $u_r := y_1$ and $y_r := y_n$,
		\item $\calG_r(\bx_r) := \calG_1(\bx_1)\land\calG_2(\bx_2) \land\dots\land\calG_n(\bx_n)$,
		\item $\calA_r := \calA_1 = \cdots = \calA_n$,
	\end{itemize}
where $"\land"$ is the conjunction operator.
\end{definition}
\begin{remark}
  Contract composition allows one to compose multiple sub-contracts of causally dependent components into one root contract. In this paper, we assume that the root contract and sub-contracts share the same assumption since the root system and all the components operate in the same environment. The assumptions for the sub-contracts are expected to be at least as general as that of the root contract. Since they cannot be stricter than the original root contract's assumptions, having the same assumptions is sufficient. For more general cases, readers can find the definition of contract composition in \cite{AG_Contracts}.
\end{remark}

In eq.~(\ref{eq:decom1}), $C_i:=(u_i, y_i, \calA_i, \calG_i(\bx_i))$ is the sub-contract for component $S_i$, and $\calG_i(\bx_i)$ is defined by
\begin{eqnarray}
  \calG_i(\bx_i) := \{ x_i\in\calX_i | x_i \le \bx_i\},  \nonumber
\end{eqnarray}
where $\bx_i\in\calX_i$ is the threshold for $x_i$.

Through contract decomposition, we can achieve rapid detection of faults by observing the failure of the sub-contracts, i.e., we can conclude that $S_i$ incurs a fault whenever $S_i$ violates $C_i$. However, according to Definition \ref{def:com} in the Appendix, we note that any violation of the sub-contract $C_i$ will lead to the failure of root contract $C_r$. Hence, simple contract decomposition will make the system $S_r$ sensitive to random disturbances and false alarms. Any false alarms from the components will shut down $S_r$. We claim that this straight decomposition has zero flexibility to resolve disturbances and false alarms.

To solve the above issue, in our previous work \cite{COMP_HContract}, we have developed a contract-based hierarchy to monitor the execution latency of the CPSs. However, the creation of the hierarchy was done manually and thus not feasible for large-scale CPS. Hence, in this paper, we will use the same approach to construct a contract-based hierarchy to monitor the NFP through algorithmic decomposition and refinement. The following definition characterizes a valid contract-based hierarchy.

\begin{definition}(Valid Contract-Based Hierarchy)\label{def:validity}
Consider a root contract $C_r$ and a group of sub-contracts $\{C_1, \dots , C_n\}$. They are said to form a valid contract-based hierarchy iff
\begin{eqnarray}
 C_1\otimes C_2 \otimes \cdots \otimes C_n \preceq C_r.  \label{eq:refine}
\end{eqnarray}
where $\preceq$ is the operator of contract refinement, defined in Definition~\ref{def:refine}.
\end{definition}
\begin{definition}
    \label{def:refine}
	{Refinement (Definition in~\cite[Table IV]{AG_Contracts}):}
	Contract $C'_i$ refines contract $C_i$, denoted by $C'_i \preceq C_i$, if and only if the following are satisfied:
	\begin{eqnarray}
	\calA_i \Rightarrow \calA'_i, \ \text{and} \ \
	\calG'_i(\bx'_i) \Rightarrow  \calG_i(\bx_i). \nonumber
	\end{eqnarray}
\end{definition}
\begin{remark}
  Contract refinement allows one to refine a contract with a weaker or the same set of assumptions and a stronger or identical set of guarantees.
\end{remark}

According to Definition~\ref{def:validity} and \ref{def:refine}, a valid contract-based hierarchy requires the following condition,
\begin{eqnarray}
  \biggl(\land_{i=1}^{n} \calG_i(\bx_i) \Rightarrow \calG_{r} (\bx_r)\biggl) \ \Leftrightarrow
   \ \sum_{i=1}^{n}\bx_i \le \bx_{r}, \label{eq:conIm}
\end{eqnarray}
where $\wedge$ is the operator of conjunction.

As shown in inequality (\ref{eq:conIm}), we can see that even when a component $S_i$ violates $C_i$, root contract $C_r$ may still remain valid. Hence, we conclude that the contract-based hierarchy provides a certain amount of flexibility for root system $S_r$ to resolve a certain disturbance caused by several components.

We can observe that the smaller $\bx_i$ is, the higher is the flexibility that $S_r$ will have. However, if $\bx_i$ is too small, then $S_i$ will easily violate $C_i$, leading to a high False Alarm Rate (FAR) and incur a significant communication cost due to the nature of having a hierarchy of components. Hence, by choosing different $\bx_i$, we note that there exists a tradeoff between flexibility and communication cost. Therefore, given the requirement, we will define a criteria to balance the flexibility and communication issues. Accordingly, we propose the problem statement as follows:

Problem Statement: Given a root system $S_{r}$ and its associated contract $C_r$, we generate a contract-based hierarchy such that $C_r$ and $\{C_1, \dots , C_n\}$:
\begin{enumerate}
\item Form a valid contract hierarchy as in Definition~\ref{def:validity};
\item Satisfy the specification of component $S_i$, for $i\in\calN$;
\item Satisfy an optimization criteria that evaluates the tradeoff between flexibility and communication cost.
\end{enumerate}

To solve the above problem, our mechanism has two steps: (1) decompose $C_r$ into $\{C'_1, \dots, C'_n\}$; (2) refine $\{C'_1, \dots, C'_n\}$ into $\{C_1, \dots, C_n\}$. Contracts $C_r$ and $\{C_1, \dots, C_n\}$ should satisfy condition (\ref{eq:refine}).

Note that a large-scale CPS can have numerous components. Hence, manually constructing a contract-based hierarchy can be error-prone and time-consuming. To address the issue, in Section~\ref{sec:2level}, we develop algorithms to generate a valid contract-based hierarchy automatically. The algorithms have two steps: decompose root contract $C_r$ based on the specifications of the components and designers' requirements; and refine the sub-contracts $\{C_i\}_{i=1}^n$ based on the proposed criteria, related to the flexibility and communication costs.

\begin{figure}[thp]
  \centering
  \includegraphics[width=6cm]{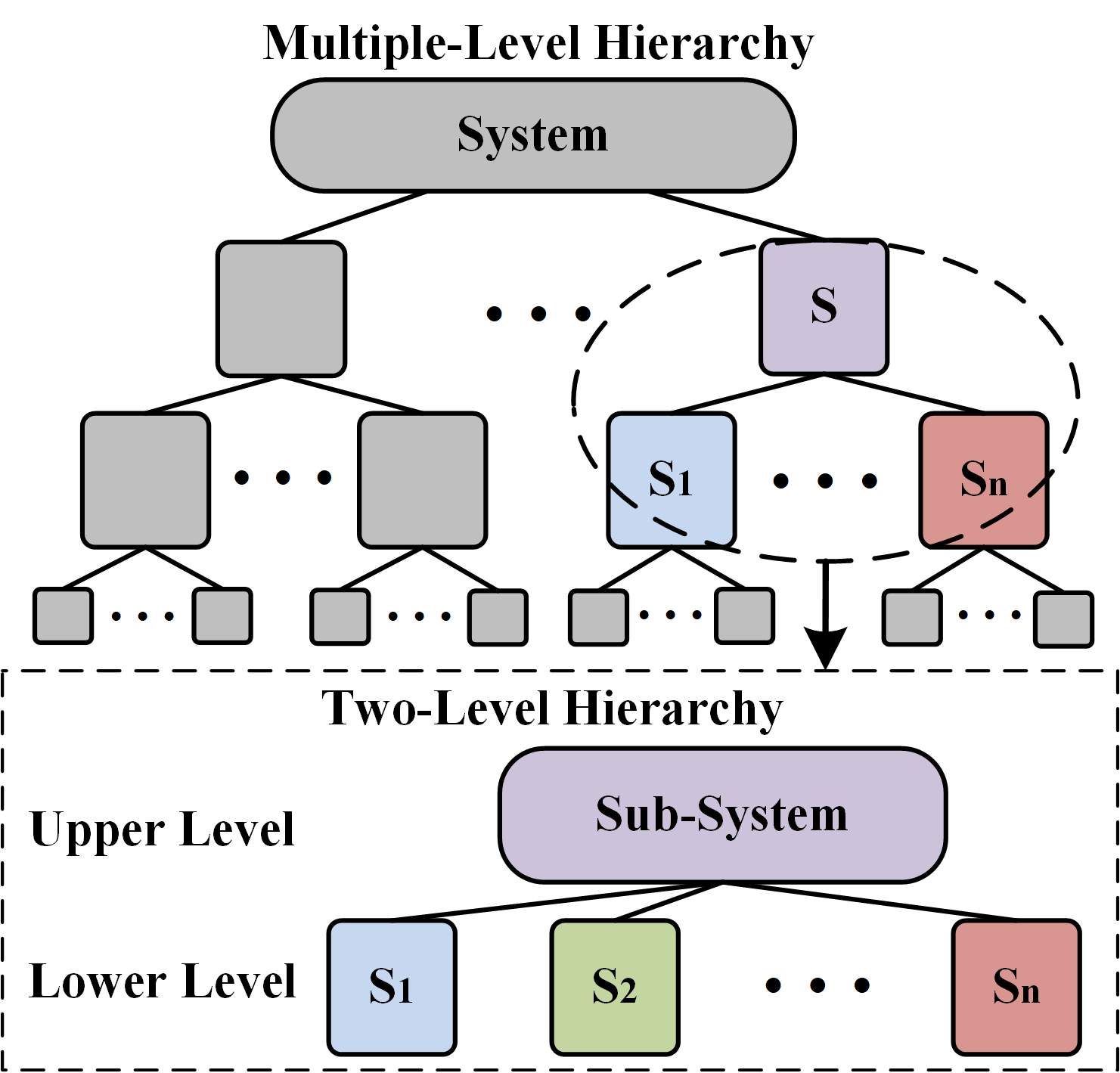}
  \caption{The Multi-Level Hierarchical Architecture of a CPS: We can view any multi-level hierarchy as a combination of multiple two-level hierarchies.}\label{fig:hierArch}
  \vspace{-2mm}
\end{figure}

If a CPS has multiple NFPs of interests, we can use the proposed mechanism to generate multiple independent contract hierarchies for monitoring different NFPs. Although our framework focuses on the generation of a two-level contract-based hierarchy, we can extend our work to multi-level hierarchies. As illustrated in Fig.~\ref{fig:hierArch}, we can view a multi-level hierarchy as a combination of multiple two-level hierarchies under the same assumptions on NFPs. To generate a multi-level hierarchy, we can run the algorithms, proposed in Section~\ref{sec:2level}, iteratively. For the rest of the paper, we will focus on the generation problem of a two-level contract-based hierarchy.

%% file: AutoGen.tex
\section{Two-level Contract-based Hierarchy Generation} \label{sec:2level}
%Need to add in decomposition.

In this section, we present our approach to automating the generation process of a two-level contract-based hierarchy through contract decomposition and refinement. We then formulate an optimization problem which refines the sub-contracts. The refinement will satisfy the condition in Definition~\ref{def:validity}. To solve the optimization problem efficiently, we use dual decomposition, which introduces a plug-and-play feature. Whenever the operators add a new component to the system, we only need to add the corresponding sub-problem, without changing any other existing sub-problems.

\subsection{Contract-Based Hierarchy Generation}

In this subsection, we describe how to facilitate sub-contract generation. Given a root contract $ C_r $, we obtain its i) inputs $u_r$, ii) outputs $y_r$, iii) assumptions $ \calA_r $, iv) guarantees $ \calG_r $ and lastly v) NFP of interest. Likewise, we gather all available components information: i) inputs $ u_i $, ii) outputs $ y_i $, and iii) estimated value $x_i$ of the NFP.

With the information gathered, we decompose the root contract by identifying a chain of dependencies (DependencyChain) among the components  $S_i$ such that the outputs $ y_i $ of a preceding component leads to the inputs $u_j $ of the next component. The search continues until a set of chained components matches the original set of inputs and outputs of the root contract. For every component involved in the chain, we formulate a sub-contract. The  assumptions are carried over from the root contract as each hierarchy of contracts has the same assumptions. Refinement of contract parameters takes place when all the sub-contracts have been established for each contract's guarantees. The mechanism ensures that the lower-level contracts meet the requirements provided in Definition~\ref{def:com}. Algorithm~\ref{algo:contractGen} illustrates this process.

%The mechanism will carry over the assumptions and guarantees from the root contract to the lower-level contracts. The assumptions and guarantees of the root and lower-level contracts will meet the requirements provided in Definition \ref{def:dec}.

\begin{algorithm}[thp]
	\caption{Two-level Contract-based Hierarchy Generation}
	\label{algo:contractGen}
	\begin{algorithmic}
	\Function{Decompose}{$ C_r $}
	
		\State {\textbf{Input:} $C_r := ( $ $u_r$, $y_r$, $\calA_r$, $ \calG_r(\bx_r) )$}
		\State {\textbf{Output:} DependencyChain}
		\For {each component $ S_j $, $ j \in\calN $}
			\If {$ (u_r == u_j )$}
				\State {Store component $ S_j $ in DependancyChain}
				\State	\Call{FindComp}{DependancyChain,  $S_j$}
			\EndIf
		\EndFor	
	\EndFunction
		
	\Function{FindComp}{DependancyChain,  $S_i$}
		\If {$ (y_r == y_i) $}
			\State \Return DependencyChain
		\EndIf
		\For {each component $ S_k $, $ k \in\calN $ }
			\If {$ (y_i == u_k) $}
				\State {Store component $ S_k $ in DependencyChain}
				\State \Call{FindComp}{DependancyChain, $ S_k $}
				\State {\textbf{break}}
			\EndIf
		\EndFor	
	\EndFunction

	\Function{FormSubcontracts}{DependancyChain}
		\State {\textbf{Input:} DependancyChain}
		\State {\textbf{Output:} SubContracts}
		\For {each component $ S_i $ in DependancyChain}
			\State {$ C_i := ($ $ u_i $, $ y_i $, $ \calA_i $, $\calG_i(\bx_i) )$ }
			\State {Where $ u_i $ and $ y_i $ are the inputs and outputs of $ S_i $ }
			\State {$ \calA_i  :=  \calA_r $ }
			\State {$ \calG_i(\bx_i)$ is assigned during \Call{Refinement}{}}
		\EndFor
		\State {\Call{Refinement}{} for $ \bx_i $ values} \Comment{Algorithm~\ref{algo:refinement}}
		\State \Return SubContracts
	\EndFunction
	\end{algorithmic}
\end{algorithm}

In the next subsection, we formulate an optimization problem to generate the parameters for the sub-contracts automatically. The optimization problem ensures that the composition of the sub-contracts will be a refinement of the root contract.

\subsection{Automatic Contract Refinement}

We can use Algorithm~\ref{algo:contractGen} to decompose root contract $C_{r}$ into multiple sub-contracts $\{C_1, \dots, C_n\}$. However, we have not determined threshold $\bx_i$ for sub-contract $C_i$. In this subsection, we will develop an algorithm to find the optimal $\bx^*_i$ while ensuring that $C_{r}$ and $\{C_1, \dots, C_n\}$ constitute a valid contract hierarchy, as presented in Definition \ref{def:validity}.

As discussed in Section II-B, we note that there exists a tradeoff between flexibility and communication cost when choosing $\bx_i$, for $i\in\calN$. To evaluate tradeoff of choosing $\bx_i\in\calX_i$, we define an overall cost function $H_i:\calX_i\rightarrow\R$, given by
\begin{eqnarray}
  H_i(\bx_i): = \theta_i Q_i(\bx_i) + (1-\theta_i) W_i(\bx_i), \
  \forall i\in\calN, \label{eq:Hfun}
\end{eqnarray}
where $Q_i:\calX_i\rightarrow\R$ is the flexibility-cost function, and $W_i:$ $\calX_i\rightarrow\R$ is the communication-cost function; $\theta_i\in(0,1)$ is a coefficient to adjust the weight between flexibility and communication costs. According to (\ref{eq:Hfun}), we can adjust the tradeoffs between flexibility and communication cost by tuning parameter $\theta$.

A smaller value of $\bx_i$ can make $S_i$ violate $C_i$ easily, creating more false alarms and leading to a higher communication cost. However, a smaller $\bx_i$ will provide higher flexibility for root system $S_r$ to resolve a disturbance since the $S_r$ has greater tolerance to determine whether the root contract is violated by evaluating the overall performance of the entire system.

According to the above analysis, we note that flexibility cost will increase with $\bx_i$, while communication cost decreases with $\bx_i$. Besides, we need to prevent the system from selecting extreme solutions, e.g., choosing zero flexibility or full flexibility. Therefore, we let $Q_i$ and $W_i$ to be exponential functions to capture the marginal effect. We define that
\begin{eqnarray}
  Q_i(\bx_i):= \exp\biggl(\frac{\bx_i - \mu_i}{\sigma_i}\biggl), \
  W_i(\bx_i):= \exp\biggl(- \frac{\bx_i - \mu_i}{\sigma_i}\biggl), \nonumber
\end{eqnarray}
where $\mu_i\in\R$ and $\sigma_i\in\R$ are the mean and standard deviation of random variable $x_i\in\calX_i$. To capture the features of various components, we use term $\frac{\bx_i-\mu_i}{\sigma_i}$ inside the exponential functions.

To refine all the sub-contracts, we consider an overall cost function $J:\prod_{i\in\calN}\calX_i\rightarrow\R$, defined by
\begin{eqnarray}
  J(\bx_1, \dots, \bx_n) := \sum_{i=1}^{n} H_i(\bx_i).
\end{eqnarray}
Given cost function $J(\cdot)$, we formulate the following optimization problem:
\begin{eqnarray}
  \min_{\bx_1, \dots, \bx_n} &&
  J(\bx_1, \dots, \bx_n), \label{eq:MainProb} \\
  \text{subject to} &&
  \sum_{i=1}^{n} \bx_i + \phi \le \bx_{r}, \label{eq:cons1} \\
  \forall i\in\calN, && \bx_i\in\calX_i=[a_i, b_i],  \label{eq:cons2}
\end{eqnarray}
where $\phi\ge0$ is a fixed guaranteed flexibility. Constraint (\ref{eq:cons1}) implies (\ref{eq:conIm}), i.e., $\{C_1, \dots, C_n\}$ and $C_r$ can form a valid hierarchy. Even when constraint (\ref{eq:cons1}) is tight, root system $S_r$ still has $\phi$ amount of flexibility if $\phi>0$.

Before solving (\ref{eq:MainProb}), we need to analyze its feasibility, i.e., existence of the solution to problem (\ref{eq:MainProb}). We propose the following theorem to characterize the feasibility of (\ref{eq:MainProb}).

\begin{proposition}\label{thm1}
Given constraints (\ref{eq:cons1}) and (\ref{eq:cons2}), problem (\ref{eq:MainProb}) admits a feasible solution if and only if
\begin{eqnarray}
  \sum_{i=1}^{n}a_i \le \bx_{r} - \phi. \label{eq:feas}
\end{eqnarray}
\end{proposition}

\begin{IEEEproof}
   We define two sets $\calD_1$ and $\calD_2$, i.e.,
  \begin{eqnarray}
    \calD_1 &:=& \{(\bx_1, \dots, \bx_n) | \
     \bx_1 +\dots+\bx_n \le \bx_{r} - \phi\},  \nonumber \\
    \calD_2 &:=& \{(\bx_1, \dots, \bx_n) | \
     \bx_i\in[a_i, b_i], i\in\calN\}.  \nonumber
  \end{eqnarray}
  We can verify that
  \begin{eqnarray}
    \calD:= \calD_1\cap\calD_2 \ne \emptyset \quad \Leftrightarrow  \quad \sum_{i=1}^{n}a_i \le \bx_{r} - \phi. \nonumber
  \end{eqnarray}
  Note that $\calD$ is the feasible set of problem (\ref{eq:MainProb}) and function $J(\cdot)$ is continuous in $(\bx_1,\dots, \bx_n)$. Besides, $\calD$ is closed and non-empty. Therefore, problem (\ref{eq:MainProb}) must have a solution. The necessity of the theorem is straightforward.
\end{IEEEproof}

\begin{remark}
  Proposition \ref{thm1} provides a sufficient and necessary condition to guarantee the feasibility of problem (\ref{eq:MainProb}). We can verify condition (\ref{eq:feas}) before we solve the problem. In general, if the system does not meet (\ref{eq:feas}), it means that the communication constraints are too rigourous. The designer can then loosen constraint (\ref{eq:cons1}) to satisfy (\ref{eq:feas}).
\end{remark}

In the following subsection, we aim to solve problem (\ref{eq:MainProb}). However it is challenging to find a close-form solution to problem (\ref{eq:MainProb}). Hence, we develop an efficient algorithm based on dual decomposition. Besides solving the problem, dual decomposition offers two advantages: firstly, dual decomposition introduces a plug-and-play feature, i.e., when the designers add new components to the root system, we only need to add corresponding sub-problems to the algorithms; secondly, whenever the system updates the parameters, we can use the dual decomposition to update the solution efficiently.

\subsection{Dual Decomposition}

To achieve dual decomposition, we need to deal with the global constraint (\ref{eq:cons1}). Hence, we introduce a Lagrangian function, i.e.,
\begin{eqnarray}
    L(\bx_1, \dots, \bx_n, \lambda)
  &:=& \sum_{i=1}^{n}  H_i(\bx_i)
   + \lambda\biggl( \sum_{i=1}^{n} \bx_i - \bx_{r}\biggl) \nonumber \\
  &=& \sum_{i=1}^{n} \big\{ \underbrace{H_i(\bx_i) + \lambda\bx_i}_{=:L_i(\bx_i, \lambda)} \big\} -\lambda \bx_{r} \nonumber \\
  &=& \sum_{i=1}^{n} L_i(\bx_i, \lambda) - \lambda \bx_{r},
  \quad \forall \bx_i\in[a_i, b_i],  \nonumber
\end{eqnarray}
where $L_i(\bx_i, \lambda)$ is the sub-Lagrangian function of component $i$, and $\lambda\in\R_{+}$ is a Lagrangian multiplier.

Note that problem (\ref{eq:MainProb}) satisfies Slater's condition, so the dual gap is zero \cite{boyd2004convex}. We can rewrite problem (\ref{eq:MainProb}) into the following:
\begin{eqnarray}
  && \max_{\lambda} \ \min_{\bx_1, \dots, \bx_n} \
 L(\bx_1, \dots, \bx_n, \lambda)  \nonumber \\
 &=&  \max_{\lambda} \ \min_{\bx_1, \dots, \bx_n} \
 \biggl\{ \sum_{i=1}^{n} L_i(\bx_i, \lambda) - \lambda \bx_{r}  \biggl\} \nonumber \\
 &=& \max_{\lambda} \ \biggl\{\sum_{i=1}^{n} \min_{\bx_i}
      L_i(\bx_i, \lambda) - \lambda \bx_{r} \biggl\},
      \quad \forall \bx_i\in[a_i, b_i]. \nonumber
\end{eqnarray}
Therefore, for each component $i\in\calN$, we formulate the following sub-problem:
\begin{eqnarray}
   \min_{\bx_i} \ L_i(\bx_i, \lambda)
   \quad \text{subject to} \ \ \bx_i\in[a_i, b_i].  \label{eq:SubProb}
\end{eqnarray}
After solving sub-problem (\ref{eq:SubProb}), we obtain the solution $\bx^*_i$. Then, we use the gradient ascent algorithm to find the optimal $\lambda$, i.e.,
\begin{eqnarray}
  \lambda^{(\tau+1)} &=& \biggl[\lambda^{(\tau)}
  + \alpha \frac{\partial L(\bx^*_1, \dots, \bx^*_n, \lambda)}
  {\partial \lambda} \biggl]_{+} \nonumber \\
  &=& \biggl[\lambda^{(\tau)}
  + \alpha\biggl(\sum_{i=1}^{n}\bx^*_i - \bx_{r}\biggl) \biggl]_{+}, \label{subPro2}
\end{eqnarray}
where $\alpha>0$ is a step size for the dual problem, and $\tau\in\Z_{++}$ is the iteration index.

Given a $\lambda\ge0$, we present the following theorem to characterize a closed-form solution to sub-problem (\ref{eq:SubProb}).

\begin{theorem}\label{thm:closed}
Given a $\lambda\in\R_{+}$, sub-problem (\ref{eq:SubProb}) has the following closed-form solution, given by
\begin{eqnarray}
  \bx^*_i = \left\{
              \begin{array}{ll}
                g_i, & \text{~if~}g_i \in (a_i, b_i);  \\
                a_i, & \text{~if~}g_i \le a_i; \\
                b_i, & \text{~if~}g_i \ge b_i; \\
              \end{array}
            \right. \label{eq:closed}
\end{eqnarray}
where $g_i$ is short for $g_i( \theta_i, \lambda)$, defined by
\begin{eqnarray}
  g_i( \theta_i, \lambda) := \mu_i +
  \sigma_i \ln\biggl\{  \frac{\sqrt{\sigma^2_i\lambda^2+4\theta_i(1-\theta_i)}
  - \sigma_i\lambda}{2\theta_i}\biggl\}. \nonumber
\end{eqnarray}
\end{theorem}

\begin{IEEEproof}
Firstly, we consider the first-order derivative of $L_i$ with respect to $\bx_i$, which yields that
\begin{eqnarray}
  \frac{\partial L_i}{\partial \bx_i}
  &=& \theta_i\frac{\partial Q_i}{\partial \bx_i}
   + (1-\theta_i)\frac{\partial Q_i}{\partial \bx_i}  + \lambda \nonumber \\
  &=& \frac{\theta_i}{\sigma_i}
  \exp\biggl(\frac{\bx_i - \mu_i}{\sigma_i}\biggl)
   - \frac{(1-\theta_i)}{\sigma_i}
   \exp\biggl(-\frac{\bx_i - \mu_i}{\sigma_i}\biggl) + \lambda. \nonumber
\end{eqnarray}
Secondly, we compute the second-order derivative of $L_i$, i.e.,
\begin{eqnarray}
  \frac{\partial^2 L_i}{\partial \bx^2_i}
  &=& \frac{\theta_i}{\sigma^2_i}\exp\biggl(\frac{\bx_i - \mu_i}{\sigma_i}\biggl)
   + \frac{(1-\theta_i)}{\sigma^2_i}
   \exp\biggl(-\frac{\bx_i - \mu_i}{\sigma_i}\biggl)  \nonumber \\
   &>& 0, \quad \forall \bx_i\in[a_i, b_i].  \nonumber
\end{eqnarray}
Hence, function $L_i(\cdot)$ is strictly convex in $\bx_i\in[a_i, b_i]$. Then, we consider the following first-order necessary condition, given by
\begin{eqnarray}
  \Rightarrow \ 0 &=&  \theta_i\exp(\tx_i)
   - (1-\theta_i)\exp(-\tx_i) + \sigma_i\lambda, \nonumber \\
   \Rightarrow \ 0 &=&  \theta_i\exp(2\tx_i)
   + \sigma_i\lambda\exp(\tx_i) - (1-\theta_i), \nonumber
\end{eqnarray}
where $\tx_i := (\bx_i - \mu_i)/\sigma_i$. Then, we obtain that
\begin{align}
  \exp(\tx_i) = \frac{-\sigma_i\lambda \pm \sqrt{\sigma_i^2\lambda^2 + 4\theta_i(1-\theta_i)}}{2\theta_i}. \nonumber
\end{align}
Removing the non-real solution, we have
\begin{eqnarray}
  \bx^*_i = \mu_i +
  \sigma_i \ln\biggl\{  \frac{\sqrt{\sigma^2_i\lambda^2+4\theta_i(1-\theta_i)}
  - \sigma_i\lambda}{2\theta_i}\biggl\}.  \nonumber
\end{eqnarray}
Combining the above solution and box constraint (\ref{eq:cons2}) yields the closed-form solution presented in (\ref{eq:closed}).
\end{IEEEproof}

\begin{remark}
  Using the results of Theorem \ref{thm:closed}, we can efficiently compute the solution to sub-problem (\ref{eq:SubProb}) based on a given $\lambda$.
\end{remark}

Besides finding the optimal solutions, we are also interested in the global constraint (\ref{eq:cons1}). When constraint (\ref{eq:cons1}) is active, we cannot significantly change the optimal solution $\bx^*_i$ by tuning the parameter $\theta_i$ since $\sum_{i=1}^n\bx_i = \bx_{r}-\phi$ is a fixed number. To this end, we study a special case, where all $\theta_i$, for $i\in\calN$, have a same value $\theta$. To avoid an active constraint (\ref{eq:cons1}), we present the following proposition to capture the relationship between $\theta$ and $\bx_{r}$.

\begin{proposition}\label{thm:global}
  Suppose that all the components choose the same value of $\theta\in(0,1)$, i.e., $\theta_i=\theta$, and $g_i(\theta, 0)\in[a_i, b_i]$. The global constraint (\ref{eq:cons1}) is inactive if and only if
  \begin{eqnarray}
  \theta \ge
  \biggl\{ \exp\biggl[\frac{2(\bx_{r} - \phi - \tmu)}{\tsigma} \biggl]+1\biggl\}^{-1}, \label{eq:globalIn}
  \end{eqnarray}
  where $\tmu:=\sum_{i=1}^{n} \mu_i$, and $\tsigma := \sum_{i=1}^{n}\sigma_i$.
\end{proposition}

\begin{IEEEproof}
Note that if constraint (\ref{eq:cons1}) is inactive, we have $\lambda=0$. To ensure (\ref{eq:cons1}) is inactive, we must have
\begin{eqnarray}
  \sum_{i=1}^{n} g_i(\theta, 0) &=&  \sum_{i=1}^{n} \biggl\{ \mu_i +
  \frac{1}{2}\sigma_i \ln \biggl(\frac{1-\theta_i}{\theta_i}\biggl) \biggl\} \nonumber \\
  &=& \tmu + \frac{1}{2} \tsigma \ln \biggl(\frac{1-\theta}{\theta}\biggl) \le \bx_{r} - \phi, \nonumber
\end{eqnarray}
which leads to
\begin{eqnarray}
  \Leftrightarrow && \theta \ge
  \biggl\{ \exp\biggl[\frac{2(\bx_{r} + \phi - \tmu)}{\tsigma} \biggl]+1\biggl\}^{-1}. \nonumber
\end{eqnarray}
This completes the proof.
\end{IEEEproof}

\begin{remark}
  Given a $\bx_{r}$, we can use Proposition \ref{thm:global} to find the feasible $\theta$, such that constraint (\ref{eq:cons1}) is inactive. If constraint (\ref{eq:cons1}) is active, it is difficult to adjust the tradeoff between communication cost and flexibility by tuning $\theta$.
\end{remark}

Given Theorem \ref{thm:closed}, we develop the following algorithm to find the optimal solution $\bx^*_i$ for each component $i$, for $i\in\calN$. We can use Algorithm~\ref{algo:refinement} to refine each $C_i$. In the next section, we will develop a software platform based on the proposed algorithms. We will use a testbed to evaluate the performance of the proposed algorithms based on different parameters, such as $\theta$ and $\bx_{r}$.

\begin{algorithm}
	\caption{Automated Contract Refinement}
	\label{algo:refinement}
	\begin{algorithmic}
        \State \emph{Step 1:} Initialize $\lambda^{(0)} = 0.$ Set $\tau=0$.
        \BState \emph{Step 2:}
        	\For {each component or sub-system $ i $, $ i\in\calN $}
				\State {Formulate sub-problem (\ref{eq:SubProb})
	             based on $\lambda^{(\tau)}$.}
	             \State {Compute $\bx^*_{i}$ using (\ref{eq:closed})}
	        \EndFor

        \BState \emph{Step 3:}
	        \State{Compute $\lambda^{(\tau+1)} =[\lambda^{(\tau)}
	        + \alpha\big(\sum_{i=1}^{n}\bx^*_i - \bx_r\big)]_{+}$}
			\If {$|\lambda^{(\tau+1}) - \lambda^{(\tau)}|> \epsilon$}
	        \State {Set $\tau = \tau+1$. \textbf{Goto} \emph{Step 2}.}
			\EndIf

        \BState \emph{Step 4:}
			\State {Use $\bx^*_i$ to refine contract $C_i$, for $i\in\calN$.}
	\end{algorithmic}
\end{algorithm}
\vspace{-2mm}

%% file: Simulation.tex
\section{Implementation and Experimental Results}\label{sec:experiment}

In this section, we first present the software platform to run the proposed algorithms. Secondly, we introduce a testbed to study the performance of the algorithms. Based on the testbed, we develop a root contract and decompose it into three sub-contracts. Algorithm~\ref{algo:refinement} is then used to refine the sub-contracts to form a valid contract-based hierarchy. The experimental results show the tradeoff between communication cost and flexibility.

\subsection{Software Implementation}
% from an AutomationML XML file. The XML file also provides information on other components of the system. The tool that parses the XML file searches for elements $i$, for $i\in\calN$, which relates to the root contract by matching the set of $ \calI_i$ and $\calO_i $ between the two.

We use a root contract to identify the users' requirements for the CPS. Root contract information and the technical architecture of the CPS such as hardware specifics and software logic can be stored in Automation Markup Language (AML), which is an open, eXtensible Markup Language (XML) based, free data format. AML format can be used to store plant-specific engineering data~\cite{AML}. Root contract information is first extracted from the AML file as described in Algorithm~\ref{algo:contractGen}. The same AML file also provides information on the components of the system. After constructing the sub-contracts from the root contract, the refinement process in Algorithm~\ref{algo:refinement} is executed for all the sub-contracts which need their parameter values to be optimized. Both the algorithms are written in Python.

Our resilience management framework is built on top of 4DIAC~\cite{4DIAC}, which is based on the IEC 61499 standard for an event-driven function block model for distributed control systems. 4DIAC's runtime environment; FORTE, runs on three Raspberry Pi 3s (RPIs) to hold both the control logic and resilience management software for each of the three components which are described in the following section.

\subsection{Testbed}
\label{testbed}

\begin{figure}[thp]
\center
	\includegraphics[width=\columnwidth]{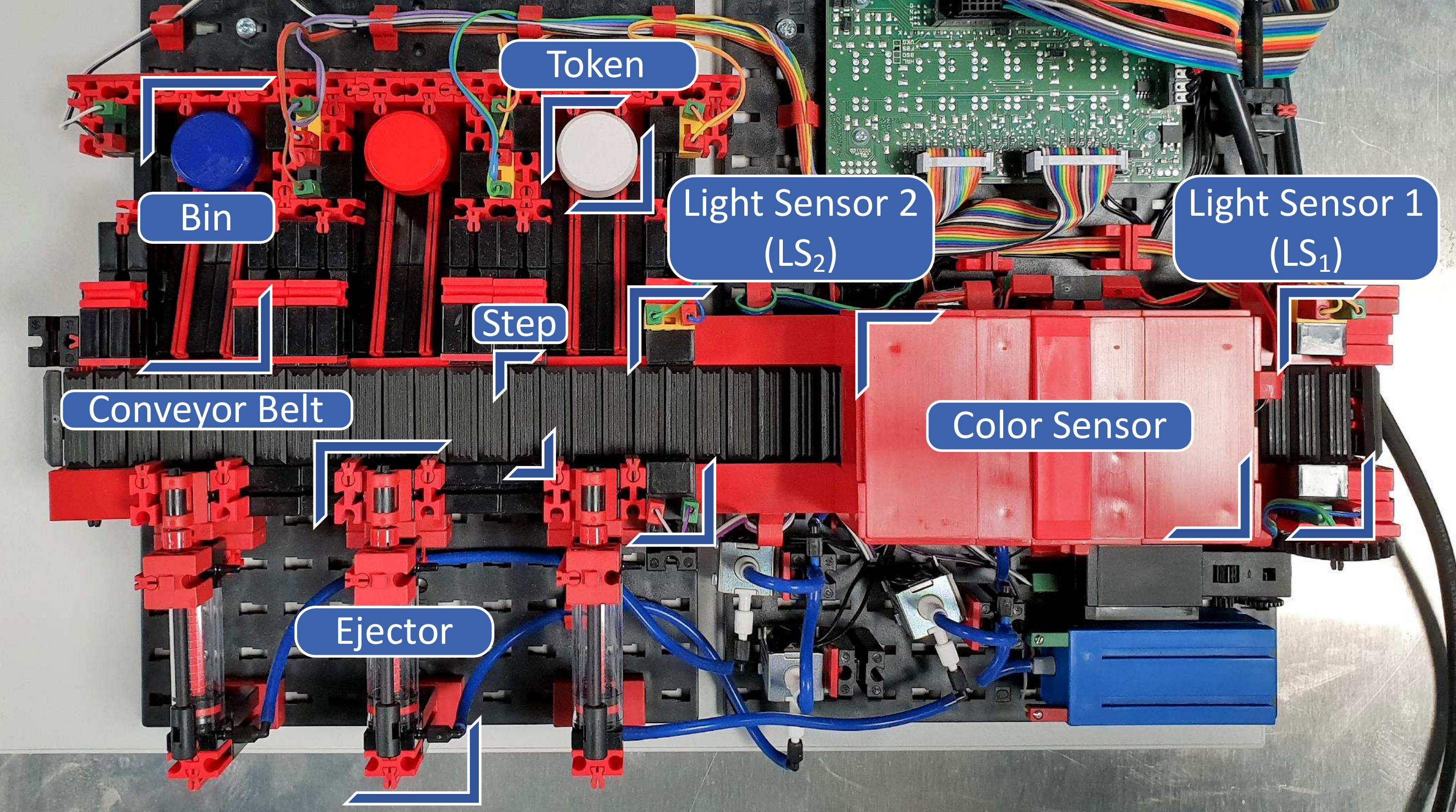}
	\caption{Fischertechnik Testbed (EAN-Code 4048962250404): It comprises of two Light Sensors (LS), a Color Processor (CP), three Ejetors (EC) and three storage bins.}
	\label{fig:Redfactory}
\end{figure}

\begin{figure}[htbp!]
\center
	\includegraphics[width=8cm]{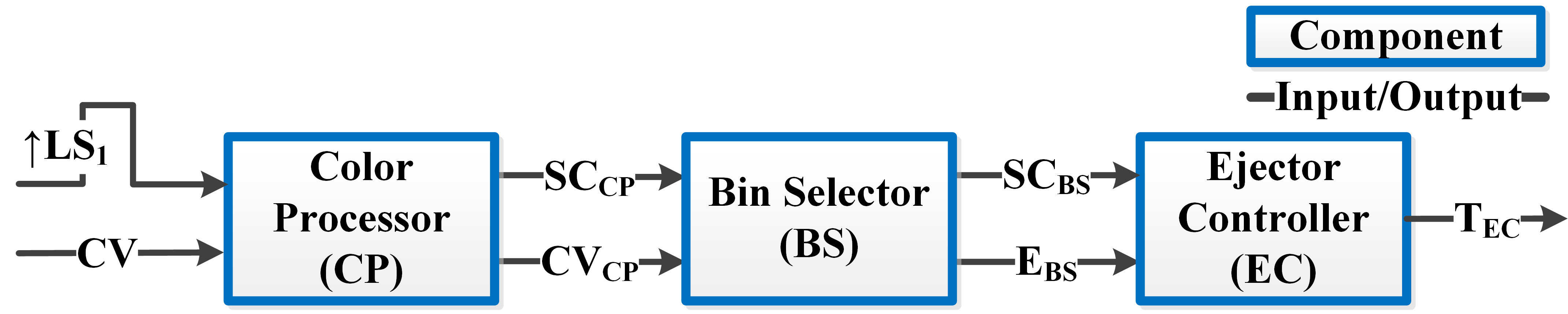}
	\caption{Functional Diagram of the Sorting Line: the CP will detect the color of the token when $LS_1$ is triggered; the CP will informs BS, which will select the right bin for the token; EC will eject the token to the bin.}
	\label{fig:FunctionalFlow}
\end{figure}

A Fischertechnik testbed, shown in Figure~\ref{fig:Redfactory}, is a CPS of a sorting line with color detection. There is a physical process of a token entering from Light Sensor 1 ($ LS_1 $), passing through the Color Processor (CP) for color identification, token sorting based on the color by the Bin Selector (BS) and eventually being ejected into the correct storage bin by the Ejector Controller (EC) as shown in Figure~\ref{fig:FunctionalFlow}. Hence, in this example, we have three key components: CP, BS, and EC (i.e., $n=3$).

\begin{figure}[thp]
	\center
	\includegraphics[width=8cm]{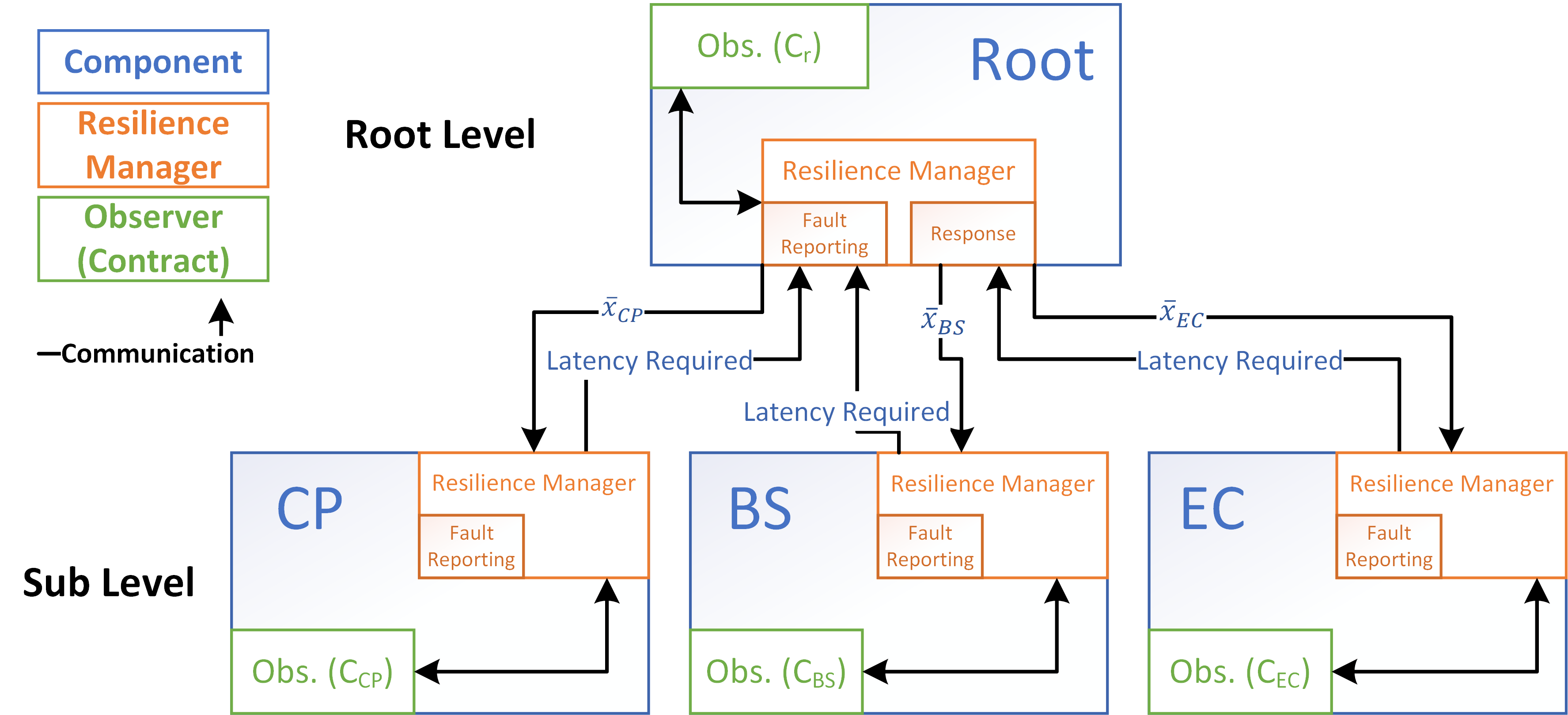}
	\caption{Two-Level Contract-Based Hierarchy: there are a root contract, which monitors the overall performance of the whole system, and three sub-contracts, which monitor the behaviors of the corresponding components.}
	\label{fig:ContractHier}
\end{figure}

In this example, the NFP is the execution latency of each component. Latency requirements for CP, BS, and EC are individually managed by their own Resilience Managers (RMs) who report to a higher level root RM. This physical flow of the token from $ LS_1 $ to the bin has an end-to-end latency requirement as the conveyor belt is always moving. The processes in between for the individual components (CP, BS, and EC) are constrained by this timing requirement and they each need to be allocated a certain latency limit for their execution. The higher-level root RM ensures that the overall end-to-end latency requirement is met. Figure~\ref{fig:ContractHier} shows the expected two-level hierarchy that would be produced by our automated tool. The root level consists of the root contract $ C_r $, the RM that is in charge of it and the observer. At the sub level, we have the three individual components, with their associated sub-contracts and they each report to the root RM.

The automated tool is used to generate the hierarchy of contracts for the sorting line from the root contract stored in the AML file. The root contract provided is the contract $C_r$ that represents the latency requirement described for the sorting line. We now present components $ S_{CP}, S_{BS} $,  $ S_{EC} $, and root contract $C_r$ formally as stated in Definition \ref{def:comp} and \ref{def:contract}, respectively. Refer to Table~\ref{table:notations} for the notations used.

\begin{contract}{Root Contract $C_r$}
	\begin{itemize}
		\item $ u_{r}: \{LS_1, CV\} $,
		\item $ y_{r}: \{T_{EC}\} $,
		\item $ \calA_{r}: \{ (M_S == s_1)\}$
		\item $ \calG_{r}(\bx_r)$: The total latency should satisfy $x_r\le\bx_r$,
	\end{itemize}
	\label{contract:root}
\end{contract}

Root contract $C_r$ handles the execution latency of the process shown in Figure~\ref{fig:FunctionalFlow}. It takes two inputs: an LS trigger $ (LS_1) $ and a color sensor value $CV\in\{$W, N, Null$\}$, where ``W" stands for white color, ``N" stands for non-white, and ``Null" means no signal. For the same motor speed $ s $, the system should generate a ejection trigger $ T_{EC} $ within $ \bx_r $.

%\begin{contract}{Root Contract $C_r$}
%	\begin{itemize}
%		\item $ \calU_{r}: \{LS_1, CV\} $,
%		\item $ \calY_{r}: \{T_{EC}\} $,
%		\item $ \calA_{r}: \{ (M_S == s_1)\}
%		\item $ \calG_{r}$: $\{\uparrow LS_1 \land CV \neq 0 \Rightarrow T_{EC}\neq\emptyset$ within $\bx_{r}$ ms\},
%	\end{itemize}
%	\label{contact:root}
%\end{contract}
%\lor (M_S == S_2) \lor (M_S == S_3) \}

%The formal specifications of CP, BS and EC are as follows:

\begin{component}{Color Processor $ S_{CP} $}
	\begin{itemize}
		\item $ u_{CP}: \{LS_1, CV\} $,
		\item $ y_{CP}: \{SC_{CP}, CV_{CP}\} $,
		\item $ x_{CP}:$ $\mu_{CP}$ and $\sigma_{CP}$
	\end{itemize}
	\label{comp:CP}
\end{component}

The Color Processor Component $ S_{CP} $ takes two inputs, $ LS_1 $ and $ CV $, and produces two outputs, $ SC_{CP} $ and $ CV_{CP} $. The component's $ x_{CP}$ characteristics of its mean $\mu_{CP}$ and standard deviation $\sigma_{CP}$ are provided based on its execution latency.

\begin{component}{Bin Selector $ S_{BS} $}
	\begin{itemize}
		\item $ u_{BS}: \{SC_{CP}, CV_{CP}\} $,
		\item $ y_{BS}: \{SC_{BS}, E_{BS}\} $,
		\item $ x_{BS}:$ $\mu_{BS}$ and $\sigma_{BS}$
	\end{itemize}
	\label{comp:BS}
\end{component}

The second component $ S_{BS} $ receives two inputs from component $ S_{CP} $: $ SC_{CP} $ and $ CV_{CP} $; and produces two outputs, $ SC_{BS} $ and bin number $E_{BS}\in\{B_1, B_2, \text{Null}\}$, where $``B_1"$ stands for Bin 1, $``B_2"$ stands for Bin 2, and ``Null" stands for no selection. The mean and standard deviation of execution time of $ S_{BS}$ are provided as well.

\begin{component}{Ejector Controller $ S_{EC} $}
	\begin{itemize}
		\item $ u_{EC}: \{SC_{BS}, E_{BS}\} $,
		\item $ y_{EC}: \{T_{EC}\} $,
		\item $ x_{EC}:$ $\mu_{EC}$ and $\sigma_{EC}$
	\end{itemize}
	\label{comp:EC}
\end{component}

Likewise, the third and last component $ S_{EC} $ receives two inputs from component $ S_{BS} $: $ SC_{BS} $ and $ E_{BS} $; and produces one output: $ T_{BS} $. It also has its mean and standard deviation.

% We plan to use the automated tool to generate the hierarchy of contracts for the sorting line. The root contract given is the contract $ C_{LM} $ which represents the latency requirement for the sorting line where $ I = \{LS_1, CV\} $, $ O = \{T_{EC}\} $, $ A = \{ (M_S == S_1) \lor (M_S == S_2) \lor (M_S == S_3) \} $, $ G = \{\uparrow LS_1 \land CV \neq 0 \implies T_{EC} \neq null \  within \  f_{LM}(M_S) \} $, and $ P = \{M_S\} $. The notations used are in Table~\ref{table:notations}. We will show how this root contract can be decomposed and refined in a later section. While this case study has only three components, our tool is scalable for $ n $ number of components.

\begin{table}[thp]
	\centering
	\caption{Notations}
	\begin{tabular}{||c c c c||}
		\hline
		Notation & Variable & Data Type & Initial Value \\ [0.25ex]
		\hline\hline
		$ M_S $ & Motor Speed & Enumerated & $ s $ \\
		$ LS_1 $ & Light Sensor 1 & Boolean & FALSE \\
		$ CV $ & Color Value & Integer & 0 \\
		$ CV_{CP} $ & Annotated Color Value & Enumerated & W, N, Null \\
		$ SC_{CP,BS} $ & Step Count & Integer & 0 \\
		$ E_{BS} $ & Ejector Number & Enumerated & B1, B2, null \\
		$ T_{EC} $ & Trigger Ejector & Boolean & FALSE \\ [0.25ex]
		\hline
	\end{tabular}
	\label{table:notations}
\end{table}

In the following subsections, we will study the properties of each component. Given the information of the components, we use Algorithm~\ref{algo:refinement} to refine the sub-contracts. We will then run the testbed to evaluate the performance of the mechanism.
\begin{figure}[thp]
  \centering
  \includegraphics[width=8cm]{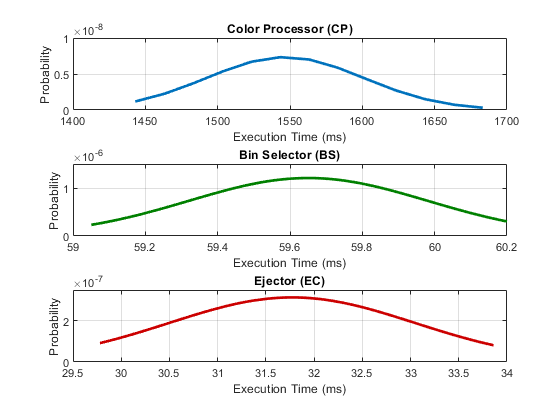}
  \caption{The empirical distributions of the components' execution latencies: all the distributions behaves like a Gaussian distribution, and different components have significantly different mean and variance of the execution latency}\label{fig:distribution}
\end{figure}

\begin{figure}[thp]
  \centering
  \includegraphics[width=8cm]{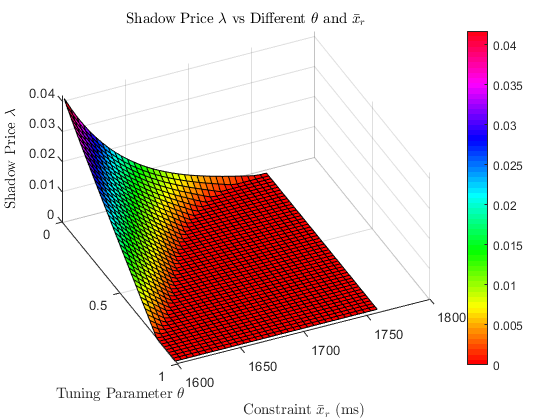}
  \caption{The relationships among parameter $\theta$, shadow price $\lambda$, and root-contract threshold $\bx_{r}$: in the red region, $\theta$ and $\bx_{r}$ satisfy condition (\ref{eq:globalIn}). However, if $\theta$ and $\bx_{r}$ do not satisfy condition (\ref{eq:globalIn}), then $\lambda$ will be positive, i.e., global constraint (\ref{eq:cons1}) stays active. }\label{fig:LamvsThetaY}
  \vspace{-2mm}
\end{figure}

\begin{figure*}
\begin{center}
\begin{minipage}[b]{0.31\linewidth}
  \centering
  \includegraphics[scale=0.38]{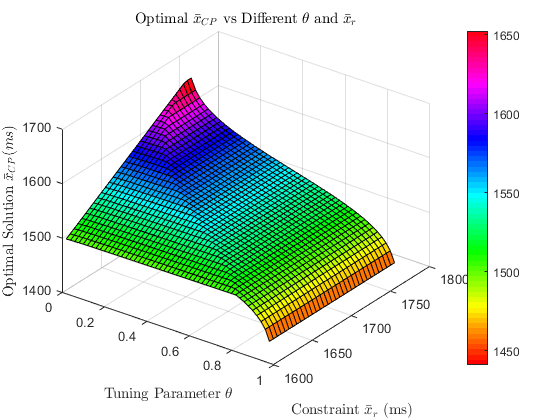}
  \caption{The relationships among $\bx^*_{CP}$, $\theta$, and $\bx_{r}$: solution $\bx^*_{CP}$ is sensitive to $\bx_{r}$. }\label{fig:xCPvsThetaY}
\end{minipage}
\hspace{0.12cm}
\begin{minipage}[b]{0.31\linewidth}
  \centering
  \includegraphics[scale=0.38]{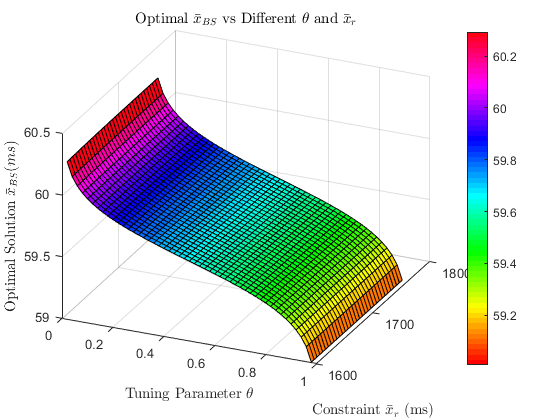}
  \caption{The relationships among $\bx^*_{BS}$, $\theta$, and $\bx_{r}$: solution $\bx^*_{BS}$ is insensitive to $\bx_{r}$.}\label{fig:xBSvsThetaY}
\end{minipage}
\hspace{0.12cm}
\begin{minipage}[b]{0.31\linewidth}
  \centering
  \includegraphics[scale=0.38]{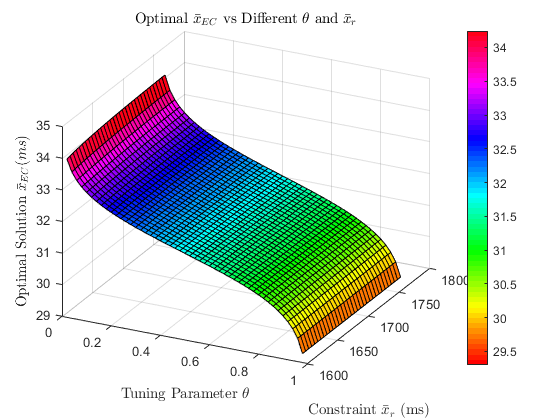}
  \caption{The relationships among $\bx^*_{EC}$, $\theta$, and $\bx_{r}$: solution $\bx^*_{EC}$ is insensitive to $\bx_{r}$.}\label{fig:xECvsThetaY}
\end{minipage}
\end{center}
\end{figure*}

\begin{figure*}
\begin{center}
\begin{minipage}[b]{0.32\linewidth}
  \centering
  \includegraphics[scale=0.38]{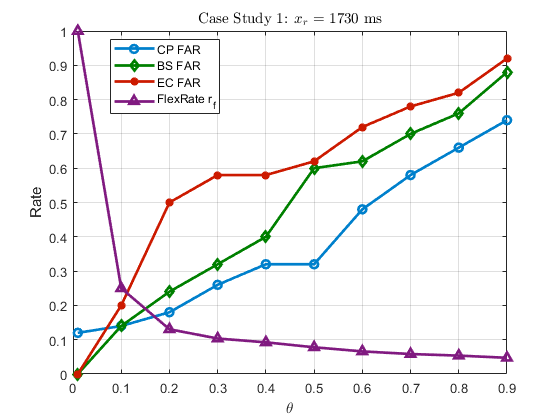}
  \caption{The false alarm rates of components and flexibility rate when $\bx_{r}=1730$ ms.}\label{fig:far1730}
\end{minipage}
\hspace{0.12cm}
\begin{minipage}[b]{0.31\linewidth}
  \centering
  \includegraphics[scale=0.38]{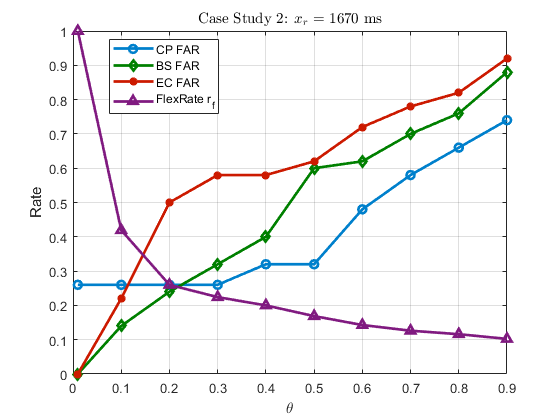}
  \caption{The false alarm rates of components and flexibility rate when $\bx_{r}=1670$ ms.}\label{fig:far1670}
\end{minipage}
\hspace{0.12cm}
\begin{minipage}[b]{0.31\linewidth}
  \centering
  \includegraphics[scale=0.38]{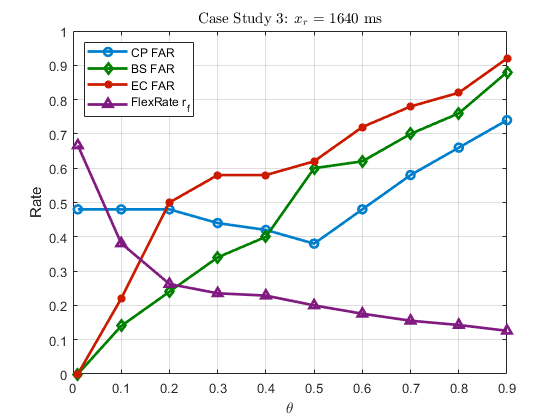}
  \caption{The false alarm rates of components and flexibility rate when $\bx_{r}=1640$ ms.}\label{fig:far1640}
\end{minipage}
\end{center}
\vspace{-4mm}
\end{figure*}

\subsection{Experimental Results}

We ran the experiments based on the capabilities of the Fischertechnik testbed. In Figure~\ref{fig:distribution}, we plot the distribution of the execution latencies of the three components based on 150 samples. We can see that each distribution behaves like the Gaussian distribution. We obtain that
\begin{eqnarray}
  \mu_{CP} &=& 1546.7 \text{~ms}, \
  \mu_{BS} = 59.651 \text{~ms}, \
  \mu_{EC} = 31.772 \text{~ms}, \nonumber \\
  \sigma_{CP} &=& 54.382 \text{~ms}, \
  \sigma_{BS} = 0.3303 \text{~ms}, \
  \sigma_{EC} = 1.2706 \text{~ms}. \nonumber
\end{eqnarray}

Given $\mu_i$ and $\sigma_i$, we use Algorithm~\ref{algo:refinement} to find the optimal $\bx^*_i$ with the same $\theta$. We study the cases where $\theta$ changes from $0.01$ to 0.99 with a resolution $0.01$, $\bx_{r}$ changes from 1600 ms to 1760 ms with a resolution 10 ms, and $\phi=10$ ms. Figure~\ref{fig:LamvsThetaY} illustrates the relationships among parameter $\theta$, shadow price $\lambda$, and root-contract threshold $\bx_{r}$. In Figure~\ref{fig:LamvsThetaY}, we can see that in the red region, the value of $\lambda$ is zero, which means constraint (\ref{eq:cons1}) is inactive. According to Proposition \ref{thm:global}, in the red region, $\theta$ and $\bx_{r}$ must satisfy condition (\ref{eq:globalIn}). In real applications, we should maintain $\lambda$ in the red region by selecting suitable $\theta$ and $x_r$. Otherwise, we cannot adjust the solution by tuning $\theta$ if constraint (\ref{eq:cons1}) is active.

Figures~\ref{fig:xCPvsThetaY}, \ref{fig:xBSvsThetaY}, and \ref{fig:xECvsThetaY} describe the relationships among tuning parameter $\theta$, root-contract threshold $\bx_{r}$, and optimal solution $\bx^*_i$ of each component. In these figures, we can see that $\bx_i$ is decreasing in $\theta$. The reason is that when $\theta$ is increasing, the weight of the flexibility cost is also increasing. Hence, the component chooses a small $\bx_i$ to increase flexibility. When $\theta$ is decreasing, the component chooses a large $\bx_i$ to reduce the communication cost. In Figure~\ref{fig:xCPvsThetaY}, when $\bx_{r}$ is sufficiently small such that condition is not satisfied, optimal solution $\bx^*_{CP}$ is insensitive to $\theta$. However, the values $\bx_{r}$ has negligible impact on $\bx^*_{BS}$ and $\bx^*_{EC}$. The reason is that the mean $\mu_{BS}$ and $\mu_{EC}$ are much smaller than $\mu_{CP}$. This feature shows that our mechanism can handle different components with heterogeneous properties with respect to the considered NFP.

In the last part, we test the tradeoff between the communication cost and flexibility. We ran the system for 50 times and sampled all the data. We create three case studies with $\bx_{r}$ = 1730 ms, 1670 ms, and 1640 ms, and in each case, $\theta$ varies from $0.01$ to $0.99$. Let $p_i$ be the FAR of component $S_i$. We define that
\begin{eqnarray}
  p_i := \frac{\text{Number of the $C_i$'s Failures}}{\text{Number of the Samples}}.
\end{eqnarray}
Since there is no real fault in the system, all the violations are false alarms. Figure~\ref{fig:far1730} shows that the FARs of the components are increasing in $\theta$. The reason is that threshold $\bx_i$ is decreasing in $\theta$ such that contract $C_i$ becomes easier to violate. Comparing to the results in Figures~\ref{fig:far1670} and \ref{fig:far1640}, we see that CP's FAR is insensitive when $\theta$ is small. The reason is that both $\theta$ and $\bx_{r}$ are small, global constraint (\ref{eq:cons1}) is active. The results indicate the limitation of the optimizer (\ref{eq:MainProb}) when condition (\ref{eq:globalIn}) is not satisfied.

Another important factor is the flexibility issues. Let $p_{r}$ be the FAR of the root contract. We define a flexibility rate as
\begin{eqnarray}
  r_f := \frac{p_{r}}{\sum_{i=1}^{n} p_{i}}. \nonumber
\end{eqnarray}
The lower the $r_f$ is, the higher the flexibility that the hierarchy will have. The reason is that a low $r_f$ means that the upper-level manager can resolve many false alarms since it has great flexibility, and vice versa. In Figures~\ref{fig:far1730}, \ref{fig:far1670}, and \ref{fig:far1640}, we see that flexibility rate $r_f$ is always decreasing in $\theta$, which coincides with our expectation that flexibility increases in $\theta$.

According to the above experimental results, we have evaluated the performance of the proposed algorithms. We have presented how to adjust the tradeoffs between communication cost and flexibility of the contract-based hierarchy. We have shown the effect of the critical region, where the optimizer, defined by (\ref{eq:MainProb}), can achieve desirable performance, such as reducing the communication cost and enhancing the flexibility. According to specific requirements, users can tune coefficient $\theta$ to design particular contract-based hierarchy, meeting the prescribed needs.

% Need to change the qCP of 10 to 9.8 

%% file: Conclusions.tex
\section{Conclusions}\label{sec:conclude}

With the growing scale of the CPSs, researchers have used contract-based technology to enhance the resiliency of the system. In our work, we have developed a contract-based hierarchy by decomposing a root contract into multiple lower-level contracts. After the decomposition, we have formulated an optimization problem to refine the parameters of the lower-level contracts, ensuring the root contract and the lower-level contract form a valid contract hierarchy. The optimization problem captures the tradeoff between communication cost and flexibility. In the experiments, we have used an application to evaluate the performance of the proposed algorithms. The results have shown that the mechanism can capture the various properties of the components as well as adjust the performance according to specific requirements. We have also analyzed the limitation of the optimizer under a given condition.

For future work, we will study the case, where the NFPs of the systems are mutually dependent. We aim to develop algorithms to construct the contract-based hierarchy. We are also interested in exploring the case, in which the NFPs satisfy the nonlinear property.